\documentclass[superscriptaddress,twocolumn,showpacs,preprintnumbers,amsmath,amssymb,prb,aps]{revtex4}
\usepackage[all]{xy}

\usepackage{graphicx,psfrag,times,epsfig,color}
\usepackage{verbatim,natbib}

\begin{document}

\title{RPA Analysis of a Two-orbital Model for the ${\rm \bf BiS_2}$-based Superconductors}
\author{G.~B. Martins} 
\email[Corresponding author: ]{martins@oakland.edu}
\affiliation{Department of Physics, Oakland University, Rochester, MI 48309, USA.}
\author{A. Moreo} 
\affiliation{Department of Physics and Astronomy, University of Tennessee, Knoxville, TN 37996 and 
Materials Science and Technology Division, Oak Ridge National Laboratory, Oak Ridge, TN 37831.}
\author{E. Dagotto} 
\affiliation{Department of Physics and Astronomy, University of Tennessee, Knoxville, TN 37996 and 
Materials Science and Technology Division, Oak Ridge National Laboratory, Oak Ridge, TN 37831.}

\begin{abstract}
{The random-phase approximation (RPA) is here applied to a two-orbital model 
for the ${\rm BiS_2}$-based superconductors that was recently proposed by Usui {\it et al.}, 
arXiv:1207.3888. Varying the density of doped electrons per Bi site, $n$, in the range 
$0.46 \leq n \leq 1.0$, the spin 
fluctuations promote competing ${A_{1g}}$ and ${B_{2g}}$ 
superconducting states with similar pairing strengths, in analogy with
the ${A_{1g}}$-${B_{1g}}$ near degeneracy found also within RPA 
in models for pnictides. 
At these band fillings, two hole-pockets centered at $(0,0)$ 
and $(\pi,\pi)$ display nearly parallel Fermi Surface segments 
close to wavevector $(\pi/2,\pi/2)$, whose distance 
increases with $n$. After introducing electronic interactions treated in the RPA,
the inter-pocket nesting of these segments leads to pair scattering 
with a rather ``local'' character in k-space. The similarity 
between the ${A_{1g}}$ and ${B_{2g}}$ channels observed here should  
manifest in experiments on ${\rm BiS_2}$-based superconductors if the pairing
is caused by spin fluctuations.
}
\end{abstract}
\maketitle

{\it Introduction.}\textemdash The recently discovered family of layered bismuth oxy-sulfide 
superconductors\cite{Mizuguchi2012b, Mizuguchi2012a, Usui2012, Li2012b, Demura2012, Tan2012a, Singh2012, Awana2012, Kotegawa2012, Zhou2012, 
Wan2012, Takatsu2012, Sathish2012, Jha2012b, Xing2012, Tan2012b, Jha2012a, Deguchi2012, Li2012a, Liu2012, Zhang2012, Lei2012} 
has immediately attracted considerable attention from the Condensed Matter community
due to its close similarities with the famous iron-pnictide 
superconductors.\cite{doi:10.1021/ja800073m,RevModPhys.83.1589,Johnston2010,Elbio2012,Dagotto2012} 
As in the case of other layered unconventional superconductors, such as the 
cuprates and the aforementioned iron pnictides/chalcogenides, this new family displays 
a layered structure involving ${\rm BiS_2}$ planes where the observed 
superconductivity is believed to reside. 
The first report of superconductivity 
originated in ${\rm Bi_4O_4S_3}$, with $T_c=4.5$ K.\cite{Mizuguchi2012b}
Superconductivity has also been reported 
in ${\it Re}{\rm O_{1-{\it x}}F_{\it x}BiS_{2}}$, 
where {\it Re} = La, Nd, Ce, and Pr, with corresponding 
$T_c=10.6$,\cite{Mizuguchi2012a} $5.6$,\cite{Demura2012} $3.0$,\cite{Xing2012} 
and $5.5$ K.\cite{Jha2012a} 
These compounds are metallic in the normal state and Density 
Functional Theory calculations indicate 
that the relevant bands crossing the Fermi surface (FS) 
originate mainly from the Bi 6$p$ orbitals, as shown, {\it e.g.}, 
for ${\rm LaO_{1-{\it x}}F_{\it x}BiS_{2}}$.\cite{Usui2012} 
However, contrary to the majority of 
the Cu- and Fe-based unconventional superconductors, no magnetically 
ordered phase has been detected thus far in the ${\rm BiS_2}$ compounds. 
This apparent absence of magnetism in the ${\rm BiS_2}$ compounds {\it may} still locate 
them in the same category as ${\rm LiFeAs}$, ${\rm FeSe}$, 
and possibly ${\rm Sr_2VO_3FeAs}$,\cite{RevModPhys.83.1589} that are also non magnetic but
their pairing properties are widely believed to still 
originate in short-range magnetic fluctuations. 
For these reasons, and despite the absence of observed long-range magnetism in ${\rm BiS_2}$, 
it is important to study the potential role of spin fluctuations in these novel materials
and the pairing channels that those fluctuations tend to favor, 
to help in the analysis of experimental data.

In this manuscript, the two-orbital (2-orbital) model recently 
introduced by Usui {\it et al.} is adopted.\cite{Usui2012}
The fact that the relevant orbitals in ${\rm BiS_2}$ compounds are 
$p$-type, where Coulomb interactions should be smaller than in $d$ orbitals, 
turns RPA into a suitable technique, whose results 
deserve a careful analysis if electron 
correlations are found to be important for superconductivity in these materials. 
Similar calculations for a related four-orbital model\cite{Usui2012} are underway. 
Note that in Ref.~\onlinecite{Usui2012} 
a brief discussion of RPA calculations has already been presented.
The results discussed by Usui 
{\it et al.} consisted of a single set of couplings 
(equivalent to our $J/U=0.2$ calculations below) at $n=0.5$. 
Their early weak-coupling RPA analysis is here expanded
via a systematic study of the influence of the band filling $n$ and the identification of
the dominant channels for superconductivity under the assumption 
of a spin fluctuations mechanism. 
The main novel contribution of our present effort is the identification
of closely competing 
${B_{2g}}$ and ${A_{1g}}$ gap functions as the dominant pairing channels, 
particularly for band fillings around $n=0.5$. 
At quarter filling ($n=1.0$), another pair of almost degenerate 
gap functions (with symmetries ${A_{2g}}$ and ${B_{1g}}$) is found to closely 
compete with the previously mentioned dominant pair, especially at $J/U=0.3$.

\begin{figure}
\centering
\begin{minipage}{1.45in}
\psfig{file=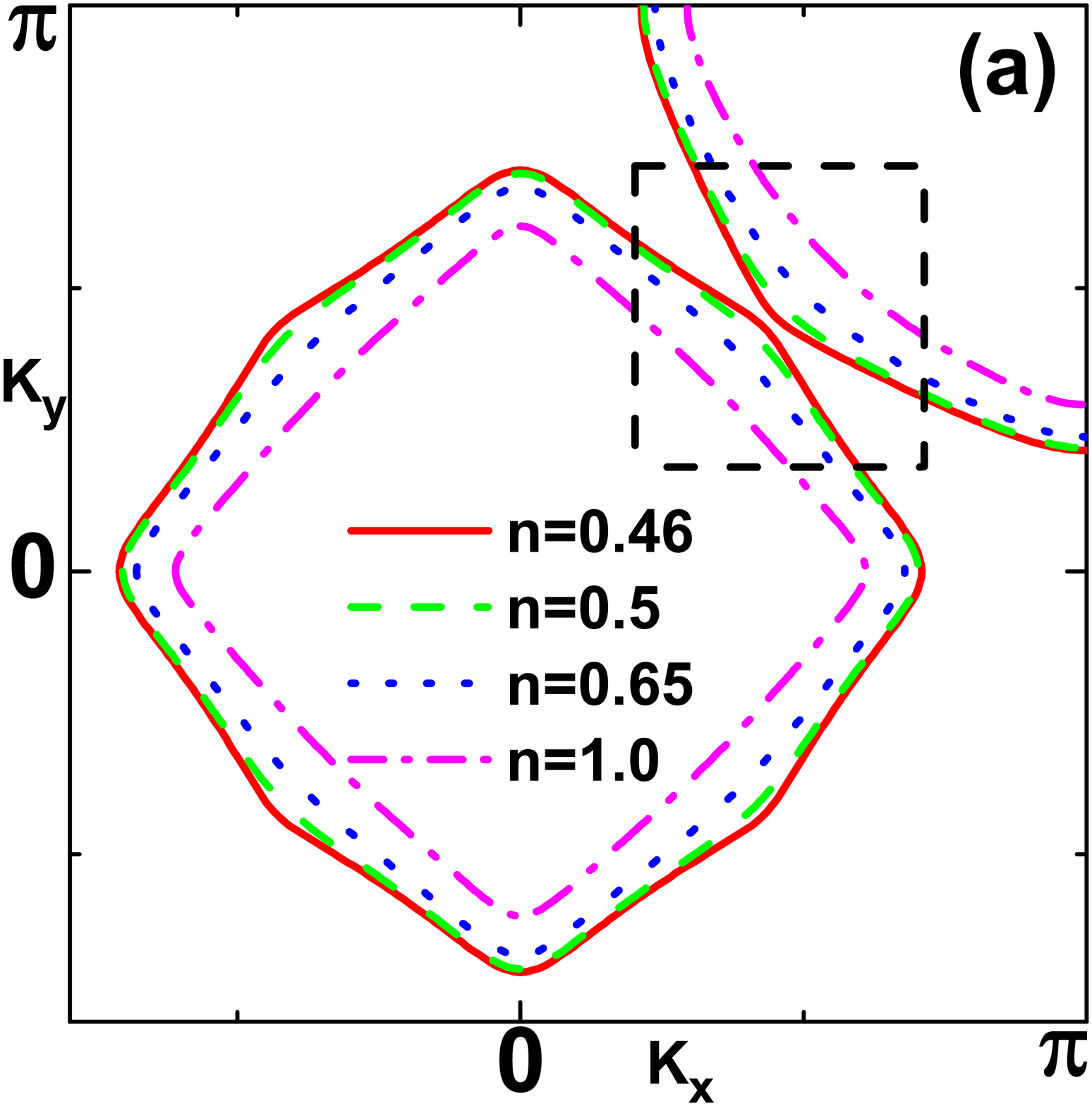,width=1.45in}
\end{minipage}
\begin{minipage}{1.75in}
\psfig{file=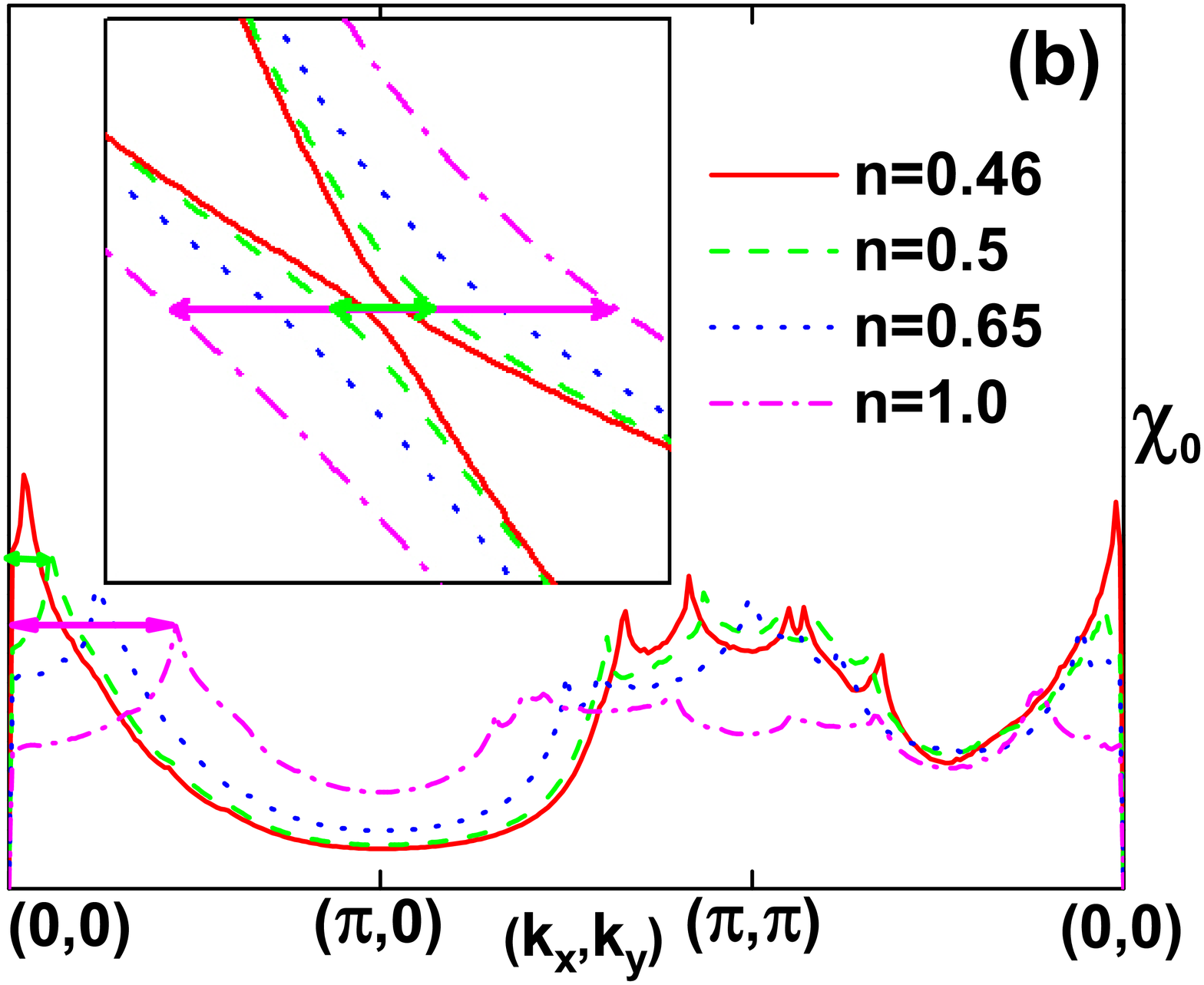,width=1.75in}
\end{minipage}
\caption{(Color online) (a) Hole-pockets for four different electronic fillings: 
$n=0.46$ (solid red), $n=0.50$ (dashed green), 
$n=0.65$ (dotted blue), and $n=1.00$ (dot-dashed magenta). 
Note that close to the $(\pi/2,\pi/2)$ wavevector, 
where the $n=0.46$ pockets almost touch, the increase of $n$ decreases the radius 
of the hole-pockets and, more importantly, the adjacent 
FS segments (inside the dashed box) become 
more and more parallel. (b) Lindhard function $\chi_0$ for the same fillings as in panel (a). 
Note that the position in k-space of the leftmost peak is 
clearly associated to FS nesting through a $(k_n,0)$ vector, as indicated in the inset, 
which zooms-in the dashed box in panel (a). Indeed, the position 
of the leftmost peaks in $\chi_0$ agree (within a few percent) with the vectors 
indicated in the inset (see text for details, especially Fig.~\ref{figure5}). 
Obviously, there are additional nesting vectors that become evident in a 2-d plot 
of $\chi_0$ [Fig.~\ref{figure5}(b)].
}
\vspace{-0.5cm}
\label{figure1}
\end{figure}

{\it Hamiltonian.} The 2-orbital model described 
by Usui {\it et al.}\cite{Usui2012} contains hopping parameters 
up to fourth neighbors, and in k-space is given by
\begin{eqnarray}\label{kspace}
H_{\rm TB}(\mathbf{ k}) &=& \sum_{\mathbf{ k},\sigma,\mu,\nu} T^{\mu\nu}
(\mathbf{ k})
d^\dagger_{\mathbf{ k},\mu,\sigma} d^{\phantom{\dagger}}_{\mathbf{ k},\nu,\sigma}~~,
\end{eqnarray}
where
\begin{eqnarray}
T^{XX} &=& 2t_x^X\left(\cos k_x +\cos k_y\right) +2t_{x \mp y}^{X} \cos  \left(k_x \pm k_y\right) \\ \nonumber 
&+& 2t_{2x \mp y}^{X}\left[\cos  \left(2k_x \pm k_y\right) + \cos  \left(k_x \pm 2k_y\right) \right]+\epsilon_X, \label{eqt11}\\
T^{YY} &=& 2t_x^Y\left(\cos k_x +\cos  k_y\right) +2t_{x \pm y}^{Y} \cos  \left(k_x \mp k_y\right) \\ \nonumber
&+& 22t_{2x \pm y}^{Y}\left[\cos  \left(2k_x \mp k_y\right) + \cos  \left(k_x \mp 2k_y\right) \right]+\epsilon_Y, \label{eqt22}\\
T^{XY} &=& T^{YX} = 2t_x^{XY}\left(\cos k_x -\cos  k_y\right) \\ \nonumber
&+& 4t_{2x}^{XY}\left(\cos 2k_x -\cos  2k_y\right) \\ \nonumber
&+& 4t_{2x+y}^{XY}\left(\cos 2k_x\cos  k_y - \cos  k_x \cos 2k_y\right) \label{eqt12}.
\end{eqnarray}
The operator $d^{\dagger}_{\mathbf{ k},\nu,\sigma}$ ($d_{\mathbf{ k},\nu,\sigma}$) 
in Eq.~(\ref{kspace}) creates (annihilates)  
an electron in band $\nu=X,Y$, with spin $\sigma=\pm$, 
and wavevector $\mathbf{ k}$. The values for the 
hopping parameters are those from Ref.~\onlinecite{Usui2012}, 
and are reproduced in Table I for completeness  (in eV units, as used throughout this paper). 
Figure \ref{figure1}(a) shows the FS hole-pockets 
for four different band fillings $n=0.46$, $0.5$, $0.65$, and $1.0$, 
with corresponding chemical potentials 
$\mu = 1.10375$, $1.12514$, $1.21828$, and $1.52621$ 
(in principle, $n=x$ in ${\rm LaO_{1-{\it x}}F_{\it x}BiS_{2}}$).\cite{Usui2012} 
Panel (b) shows the corresponding non-interacting magnetic susceptibilities $\chi_0$. 
The leftmost peaks in $\chi_0$, located at $(k_n,0)$, with $0 \lesssim k_n \lesssim \pi/2$ 
as the filling varies from $n=0.46$ to $1.0$, can be associated to FS nesting once it is noticed 
that their position matches the {\it horizontal} separation between the two adjacent FS segments 
from the pockets centered at $(0,0)$ ($\Gamma$) and $(\pi,\pi)$ ($M$), as highlighted by the dashed box 
in panel (a) and sketched in the inset to panel (b). 
Note that the {\it horizontal} separation is well defined if the two FS segments are parallel, 
which is the limiting case as $n$ increases, as shown in the inset, 
to $n=1.0$ (for details, see Fig.~\ref{figure5} and the associated 
discussion). It is also important to remark that 
once interactions are introduced, the leftmost peak in $\chi_0$ is the one that diverges 
in the RPA calculation of the {\it spin} susceptibility $\chi_{\rm RPA}$ for almost all the 
fillings and various values of interaction parameters. 
This divergence indicates a tendency to magnetic order, or at least 
strong spin fluctuations (paramagnons), 
with characteristic wavelength determined by $(k_n,0)$. 
Our analysis is not extended into the $n\leq0.45$ region 
since there the topology of the FS changes 
(see Ref.~\onlinecite{Usui2012} for details of the FS 
at lower fillings \cite{note0}). 

\begin{table}
\caption{Tight-binding parameters (eV) for 2-orbital model.\label{tab:hopp3}}
 \begin{tabular}{|ccccccccc|}\hline
$\epsilon_{X,Y}$ & $t_x^{X,Y}$ & $t_{x \mp y}^{X,Y}$ & $t_{x \pm y}^{X,Y}$ & $t_{2x \mp y}^{X,Y}$ & $t_{2x \pm y}^{X,Y}$ & $t_x^{XY}$ & $t_{2x}^{XY}$ & $t_{2x+y}^{XY}$\\
\hline
  $2.811$   & $-0.167$ & $0.880$  & $0.094$ & $0.069$ & $0.014$ & $0.107$ & $-0.028$ & $0.020$ \\ 
\hline
 \end{tabular}
\end{table}

The Coulomb interaction in the Hamiltonian is given by 
\begin{equation}\begin{split}  
  H_{\rm int}& =
  U\sum_{{\bf i},\alpha}n_{{\bf i},\alpha,\uparrow}n_{{\bf i},
    \alpha,\downarrow}
  +(U'-J/2)\sum_{{\bf i}, 
    \alpha < \beta}n_{{\bf i},\alpha}n_{{\bf i},\beta}\\
  &\quad -2J\sum_{{\bf i},\alpha < \beta}{\bf S}_{\bf{i},\alpha}\cdot{\bf S}_{\bf{i},\beta}\\
  &\quad +J\sum_{{\bf i},\alpha < \beta}(d^{\dagger}_{{\bf i},\alpha,\uparrow}
  d^{\dagger}_{{\bf i},\alpha,\downarrow}d^{\phantom{\dagger}}_{{\bf i},\beta,\downarrow}
  d^{\phantom{\dagger}}_{{\bf i},\beta,\uparrow}+h.c.),
\label{eq:Hcoul}
\end{split}\end{equation}
where the notation is standard and the many terms 
have been described elsewhere.\cite{PhysRevB.82.104508}
Here, the usual relation $U^{\prime} = U - 2J$ is assumed, and $J/U$ is a parameter. 
Calculations 
were done for $0.1 \leq J/U \leq 0.4$, in steps of $0.1$, 
for the four fillings $n=0.46$, $0.50$, $0.65$, and $1.0$. 
The multi-orbital RPA calculations performed here follow 
closely those described in Ref.~\onlinecite{Graser2009}, 
and previous works by the authors.\cite{PhysRevB.82.104508,PhysRevB.84.094519} 
All results were obtained at temperature $T=10^{-4}$ and an imaginary part 
$\eta=10^{-5}$ was used to regularize the Green's functions.

\begin{figure}
\centering
\begin{minipage}{1.65in}
\psfig{file=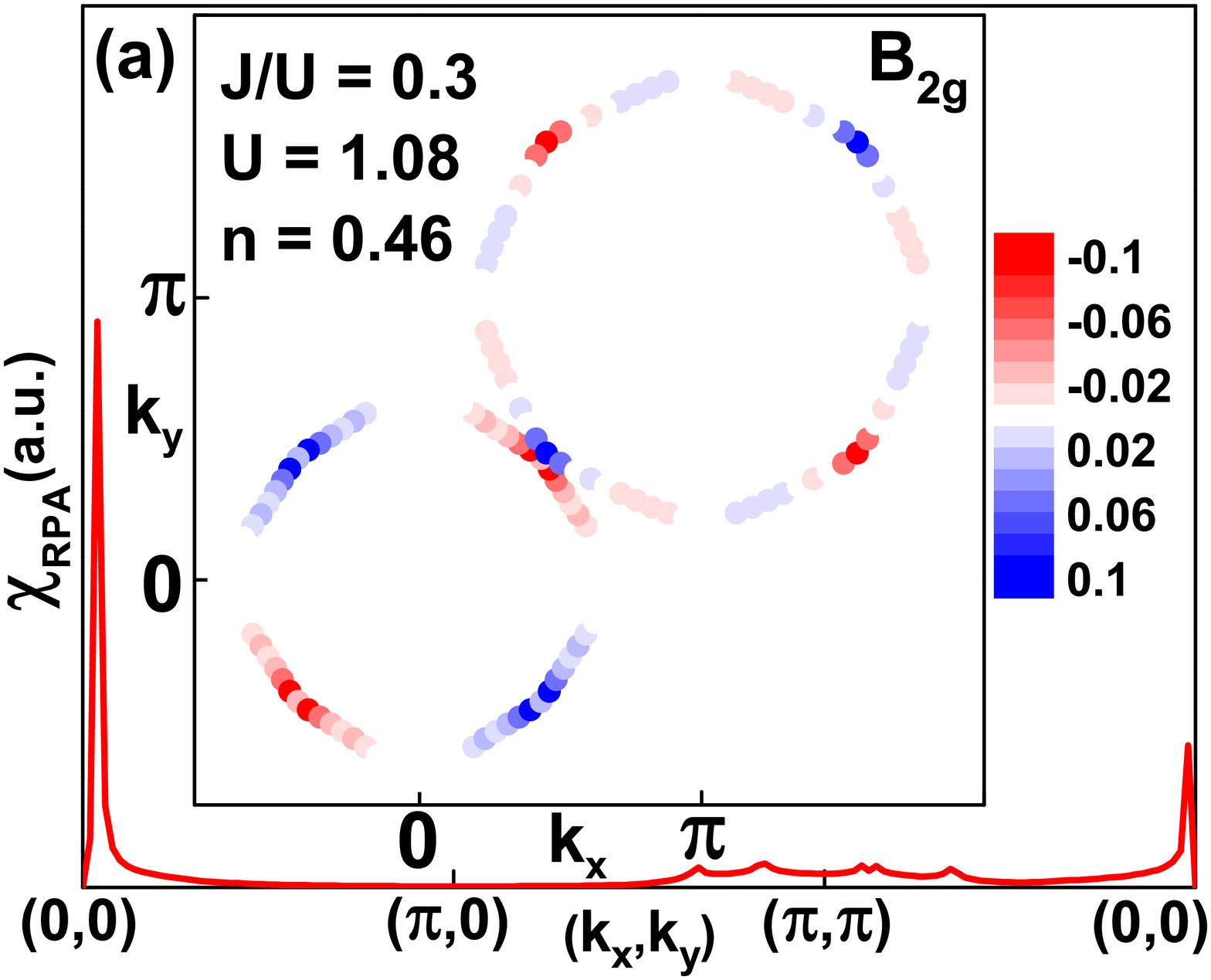,width=1.65in}
\end{minipage}
\begin{minipage}{1.65in}
\psfig{file=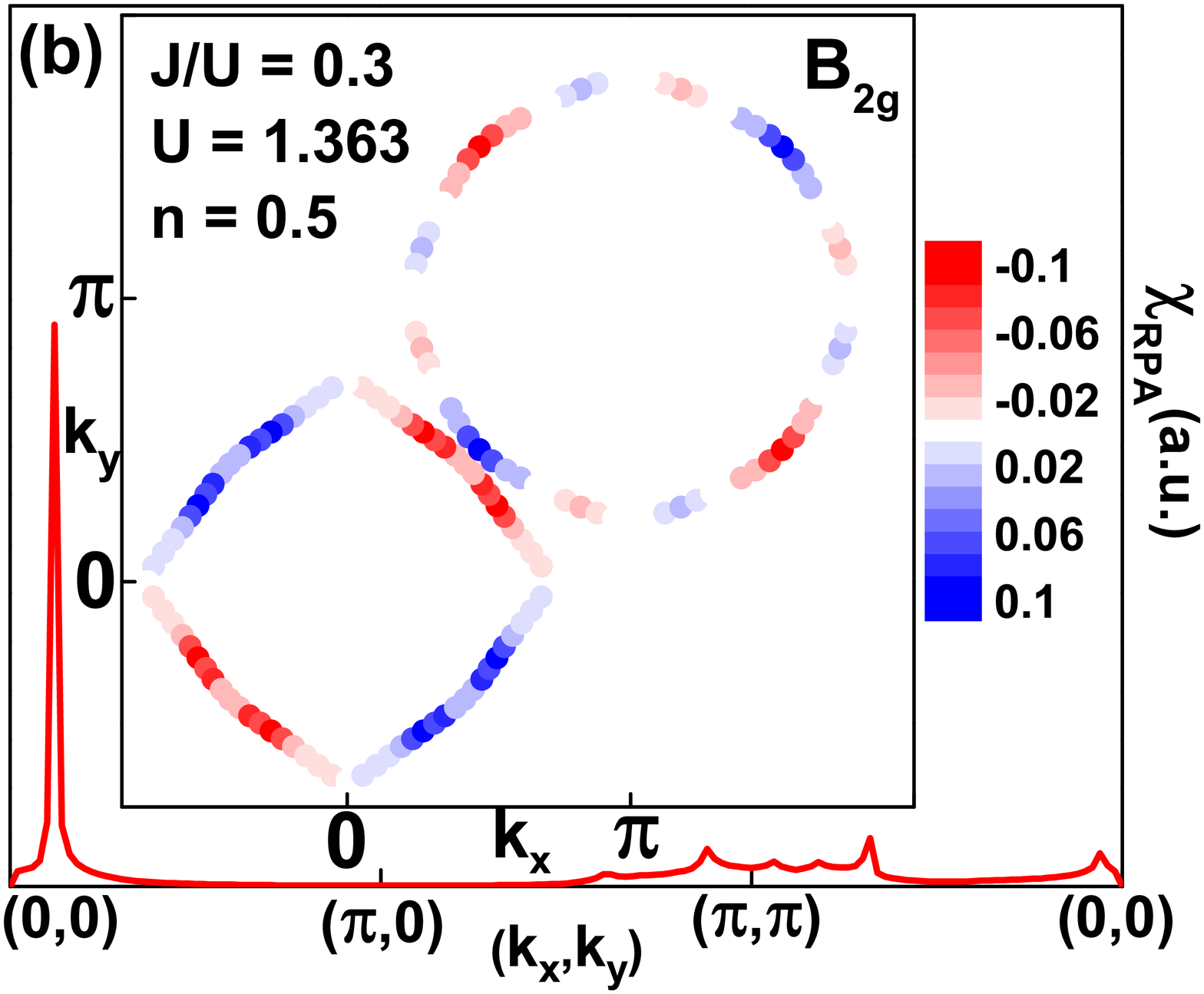,width=1.65in}
\end{minipage}
\caption{(Color online) RPA spin susceptibility (solid red curves 
in the main panels) and dominant gap function 
(red and blue dots in the insets) for (a) 
$n=0.46$ and (b) $n=0.50$. In the inset to each panel, the dominant 
gap function with symmetry ${B_{2g}}$ 
is shown. The subdominant gap function (not shown) has symmetry $A_{1g}$ and 
its eigenvalue is almost degenerate with the dominant one 
(see text).
}
\label{figure2}
\end{figure}

\begin{figure}
\centering
\begin{minipage}{1.65in}
\psfig{file=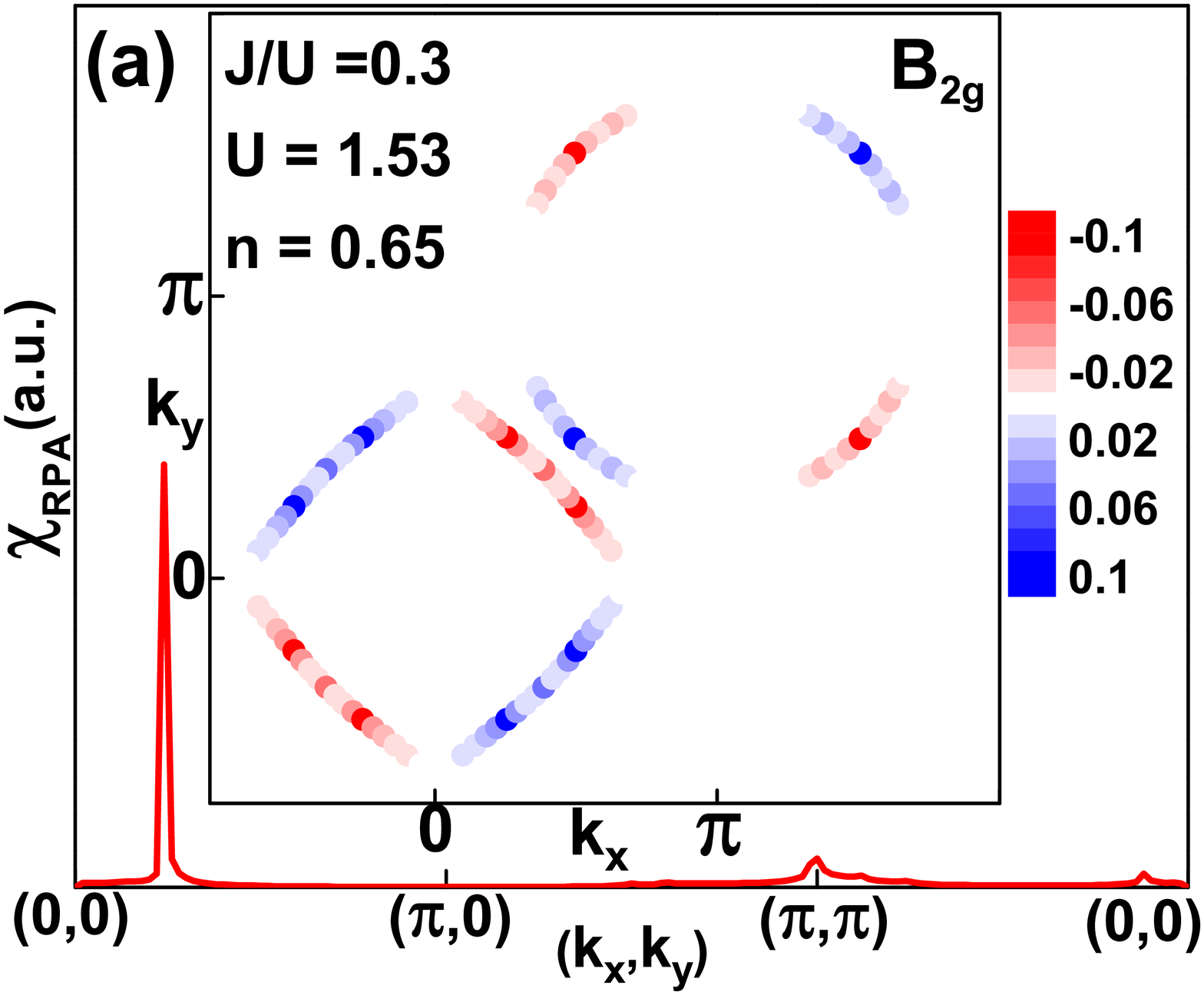,width=1.65in}
\end{minipage}
\begin{minipage}{1.65in}
\psfig{file=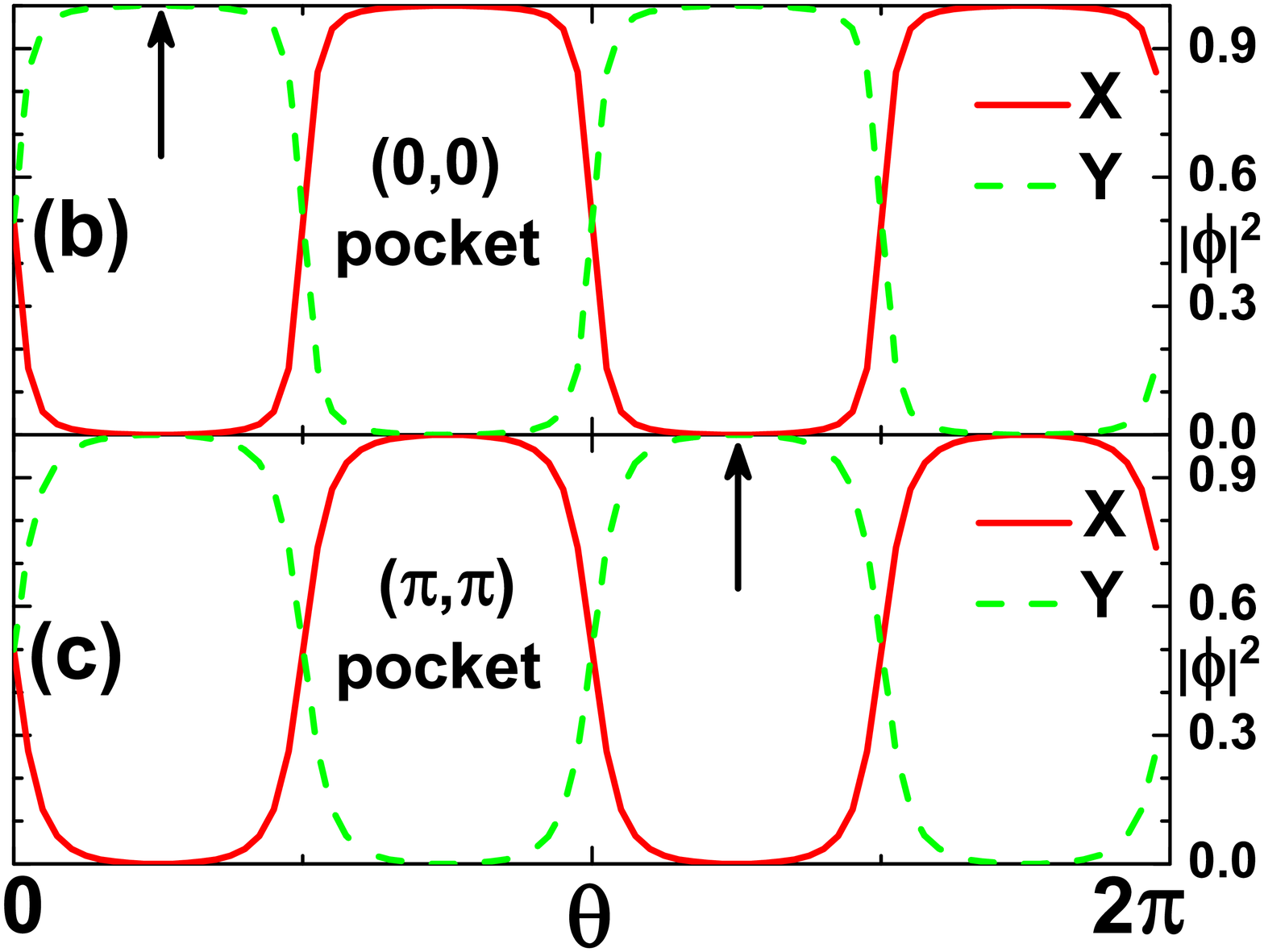,width=1.65in}
\end{minipage}
\caption{(Color online) (a) RPA spin susceptibility and dominant gap function for $n=0.65$. 
Orbital composition for the $(0,0)$ and $(\pi,\pi)$ FS pockets ($n=0.65$), 
(c) and (d), respectively. The winding angle $\theta$ is counter-clockwise, 
starting from the $k_x$ direction. Assuming the nesting described in the 
inset to Fig.~\ref{figure1}(b) as 
producing the spin fluctuations that provide pairing, 
the pair coupling is then intraorbital.
}
\label{figure3}
\end{figure}

\begin{figure}
\centering
\begin{minipage}{1.4in}
\psfig{file=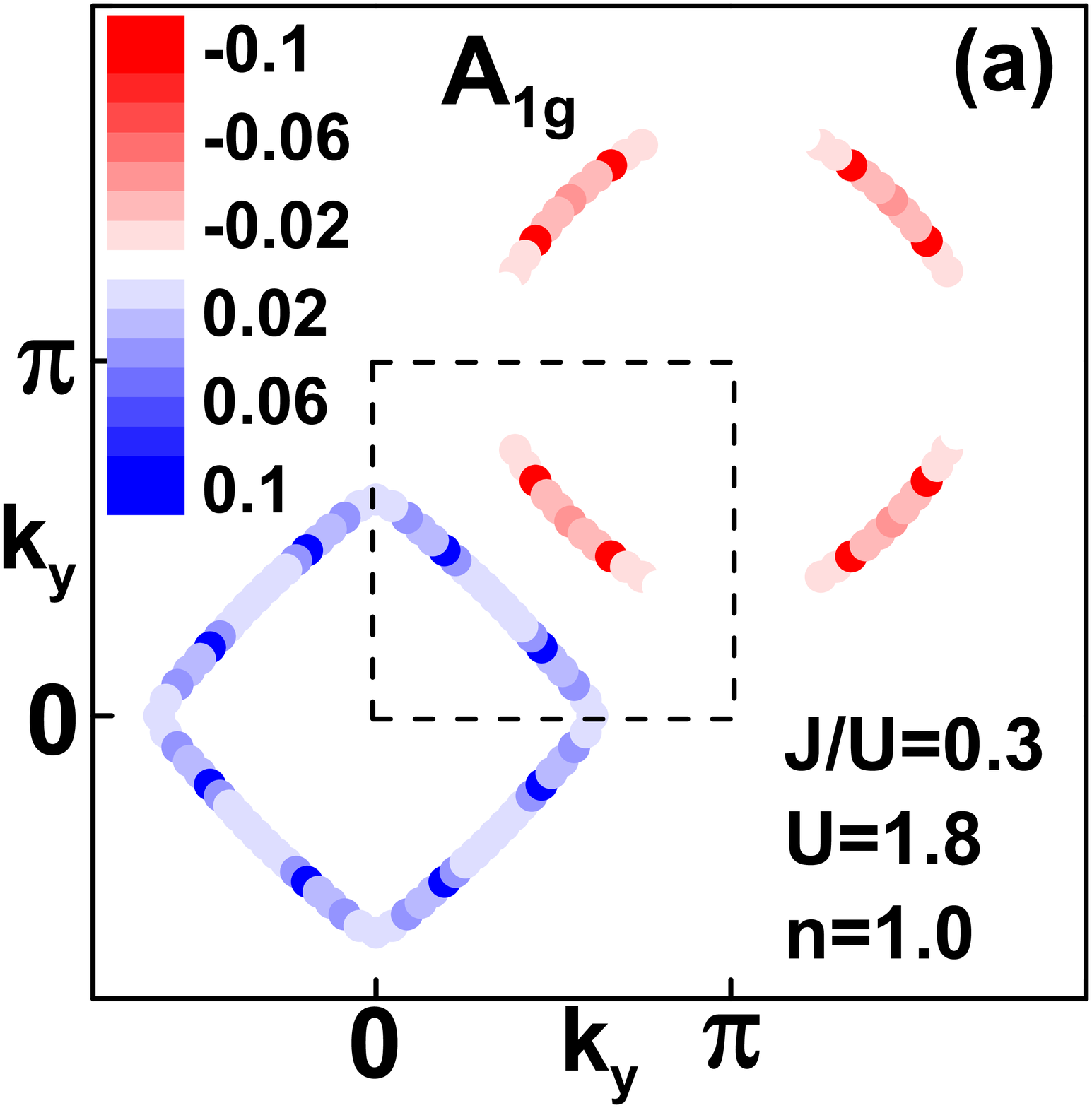,width=1.4in}
\end{minipage}
\begin{minipage}{1.8in}
\psfig{file=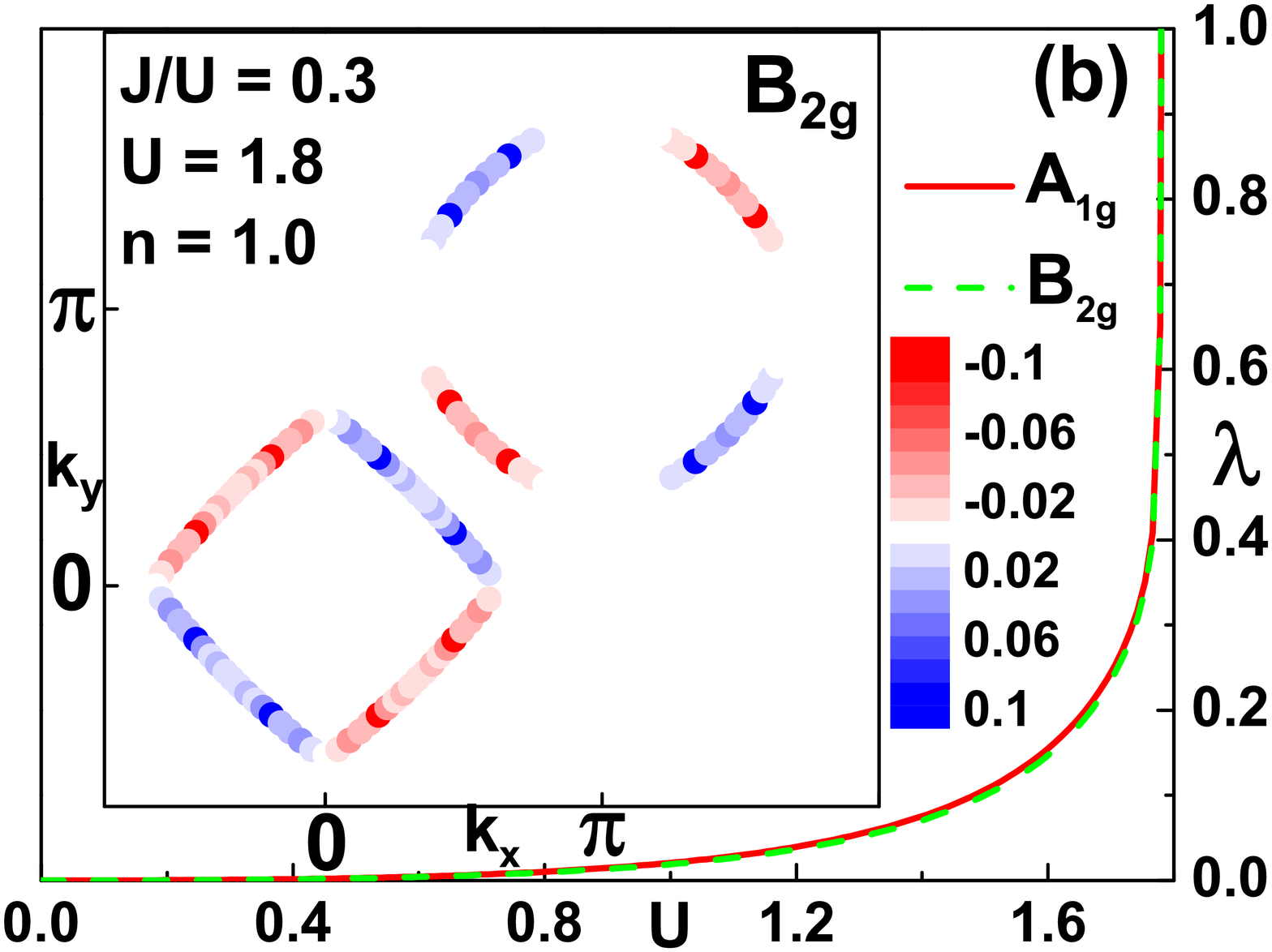,width=1.8in}
\end{minipage}
\caption{(Color online) (a) Dominant gap function with symmetry ${A_{1g}}$ at $n=1.0$.  
(b) Main panel: normalized pairing strengths $\lambda$ for the 
dominant (${A_{1g}}$, solid red curve) and 
subdominant (${B_{2g}}$, dashed green curve) gap functions. 
Although the two curves are very close, the eigenvalues are {\it not} degenerate. 
In the inset, the structure of the subdominant gap function (${B_{2g}}$) is shown. When 
compared to that of the dominant one [${A_{1g}}$ in panel (a)], 
it is clear that the structure around $(\pi/2,\pi/2)$ 
is very similar for both of them, explaining 
why the pairing strengths (eigenvalues) are the same. The region inside 
the dashed box, in panel (a), is analyzed in detail in Fig.~\ref{figure5}(a).
}
\label{figure4}
\end{figure}

\begin{figure}
\centering
\begin{minipage}{1.6in}
\psfig{file=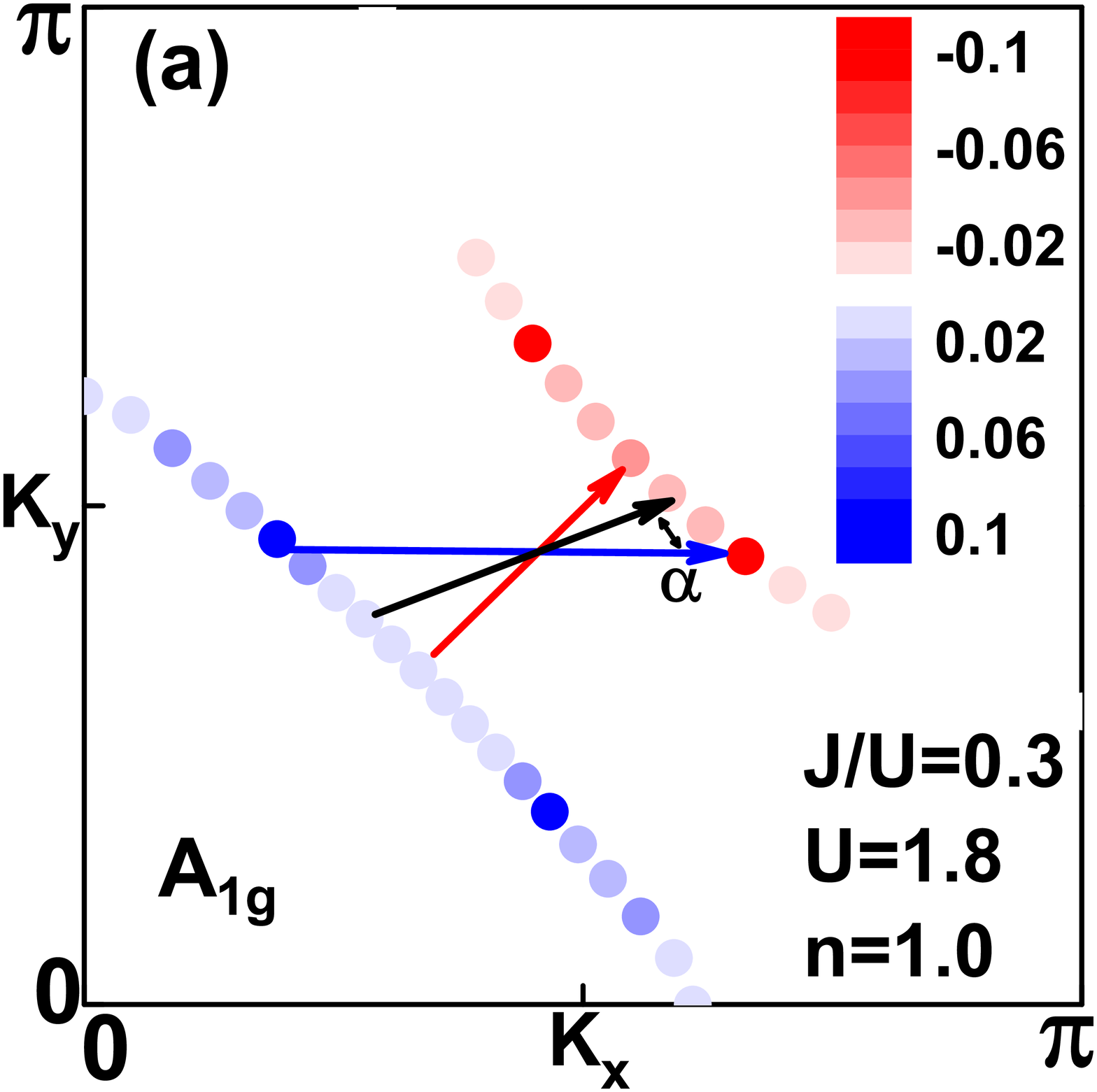,width=1.6in}
\end{minipage}
\begin{minipage}{1.6in}
\psfig{file=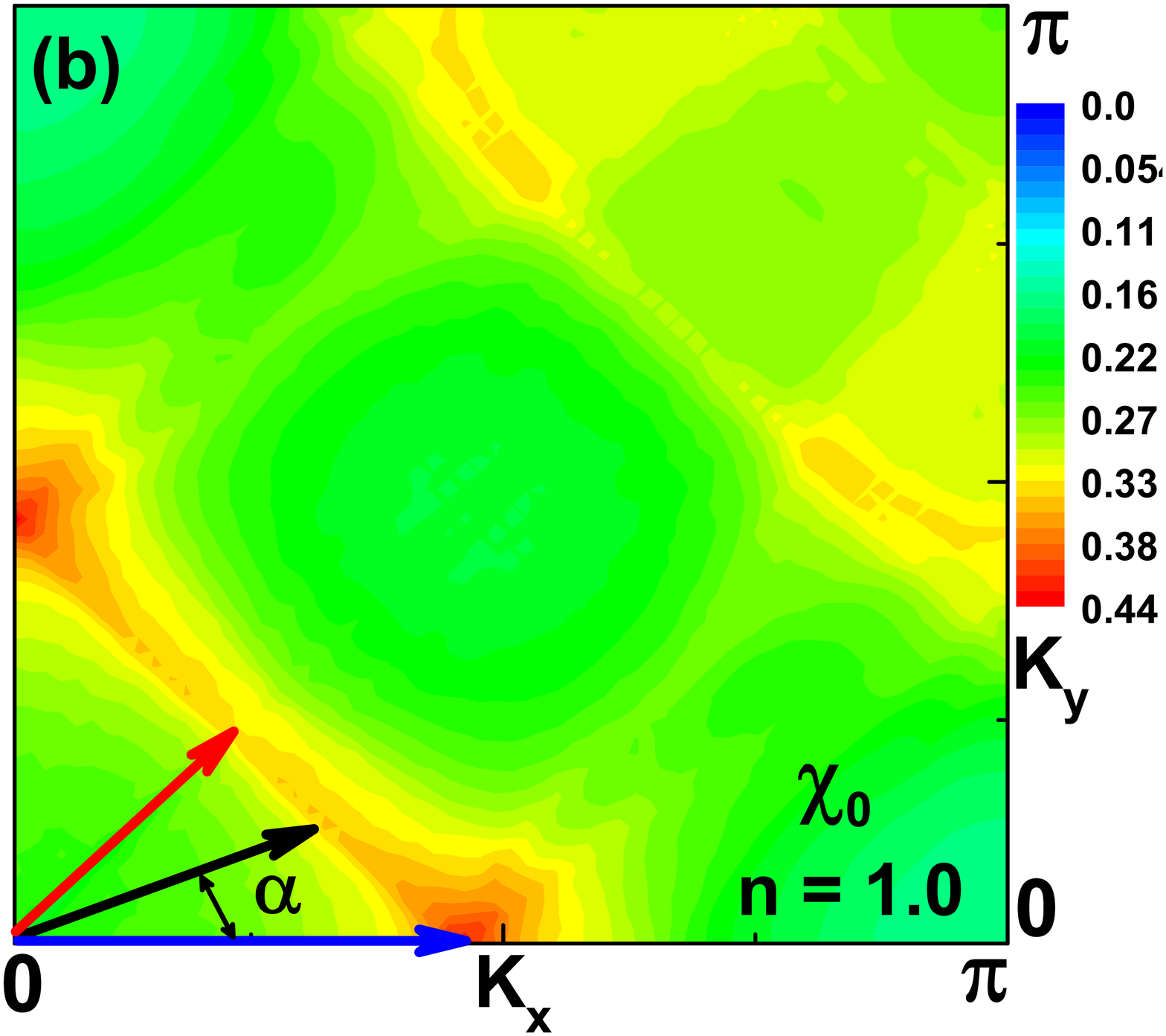,width=1.6in}
\end{minipage}
\caption{(Color online) (a) Region around point $(\pi/2,\pi/2)$ of the BZ [dashed box 
in Fig.~\ref{figure4}(a)], showing the 
dominant (${A_{1g}}$) gap-function for $n=1.0$ and $J/U=0.3$. 
(b) Two dimensional contour plot of $\chi_0$ also for $n=1.0$. The 
horizontal (blue) vector in panel (a) connects the maximum amplitude of the gap-function 
in {\it both} pockets. Note also the horizontal (blue) 
vector in panel (b) along the $k_x$ direction, 
indicating the position of the maximum value of $\chi_0$. These two vectors agree 
up to a difference smaller than the width of this maximum peak in $\chi_0$. 
Therefore, it can be shown (see text) that the line describing the position 
of the points in the $M$ pocket in relation to the points 
in the $\Gamma$ pocket, as indicated by the two additional vectors (black 
and red) in panel (a), satisfies $k_y \sim -k_x + k_n$, 
where $(k_n,0)$ and $(0,k_n)$ are the positions of the maxima in $\chi_0$ (with $n=1.0$). 
This equation also describes the line of local maxima of $\chi_0$, 
as seen in panel (b), originating from FS nesting. 
}
\label{figure5}
\end{figure}

Our RPA results for spin-singlet pairing 
link the dominant superconducting gap functions to spin fluctuations, which originate in 
FS nesting and are enhanced by electronic interactions. 
The particular relative topology of the two adjacent 
hole-pockets (see Fig.~\ref{figure1}) promotes pairing 
whose strength is independent of the global symmetry 
of the pairing functions [see Fig.~\ref{figure4}(b)]. 
Indeed, the ${B_{2g}}$ 
and ${A_{1g}}$ symmetries have essentially the 
same pairing strength, which is determined by pair scattering between 
these two adjacent FS segments (see Fig.~\ref{figure5}) 
close to $(\pi/2,\pi/2)$ in the Brillouin Zone (BZ). 
In addition, our results show that both dominant 
gap functions change sign between these two segments 
(Figs.~\ref{figure2} to \ref{figure4}), 
and the pairing is through the intraorbital 
scattering channel [Fig.~\ref{figure3}(b)]. 
The near degeneracy ${A_{1g}}$-${B_{2g}}$ is 
the analog of the near degeneracy ${A_{1g}}$-${B_{1g}}$ 
found also in RPA calculations for the pnictides,\cite{Graser2009} since
the pocket structures in both cases can be related by a 45$^{\rm o}$ rotation.
Results for spin-triplet pairing 
are presented in the supplemental material at the end of the manuscript. 

{\it Results and Discussion.} 
As mentioned above, the most important feature of the FS for fillings 
between $0.46$ and $1.0$ is that the hole-pockets centered 
at the $\Gamma$ and $M$ points present almost parallel 
segments close to the $(\pi/2,\pi/2)$ wavevector, 
becoming more and more parallel as the pockets shrink, with increasing filling 
[see Fig.~\ref{figure1}(a) and inset in Fig.~\ref{figure1}(b)]. 
In Figs.~\ref{figure2} and \ref{figure3} it will be shown 
that this has important consequences for the 
spin fluctuations and the superconducting pairing associated to this 2-orbital model. 
Indeed, as displayed in the main panel of Fig.~\ref{figure2}(a) (solid (red) curve), there is a divergence in the RPA 
spin susceptibility for very small 
$\mathbf{ k}$ values: $\mathbf{ k}_{0.46} \sim (\pi/25,0)$ for $n=0.46$, 
and  $\mathbf{ k}_{0.50} \sim (\pi/8,0)$ for $n=0.50$ [panel (b)]. A divergence 
in the spin susceptibility $\chi_{\rm RPA}$ may point to magnetic order, 
or at least to strong spin fluctuations 
with wave vector $\mathbf{ k}_n$. 
Figure \ref{figure3}(a) shows the same calculations, but now for 
$n=0.65$. Note that although $\chi_0$ 
displays a broad-peak structure around $(\pi,\pi)$ [see Fig.~\ref{figure1}(b)], 
$\chi_{\rm RPA}$ does not
present a divergence in this region. In the 
insets to Figs.~\ref{figure2}(a) and (b), and Fig.~\ref{figure3}(a), it is shown that
the dominant gap function at the FS has symmetry ${B_{2g}}$ for the three cases, 
showing that despite the changes in the size of the hole-pockets 
the results are qualitatively the same. Figures~\ref{figure3}(b) and (c) 
contain the orbital contribution ($X$, red solid curve; $Y$, green dashed curve) of the BZ 
states at the FS for the $\Gamma$ and $M$ pockets, respectively. 
It is interesting to note that the 
modifications in the position of the peak in $\chi_{\rm RPA}$ 
correlates well with the ``separation'' between the $\Gamma$ 
and $M$ hole-pockets in the region around 
$(\pi/2,\pi/2)$. For the purposes of describing 
our results, this separation will be defined as the 
{\it horizontal} distance between two parallel lines tangent to the hole-pockets at the 
points where each intercepts the $\Gamma - M$ ($\Sigma$) line. 
As described in more detail in Fig.\ref{figure5}(a) 
[and already mentioned in connection with Fig.\ref{figure1}(a)], as the filling increases 
these segments of FS approach more and more the parallel lines just defined, justifying the 
definition just given. 

The RPA results for the gap functions also point to an interesting effect, 
namely, the small value of $\mathbf{ k}_n$ for fillings $0.46 \leq n \leq 1.0$ results 
in the pairing strength depending on very ``local'' properties of the gap function  
at the adjacent segments of the hole-pockets. This implies that the pairing 
strength of gap functions with different symmetries is very similar, as long 
as they have the same ``local'' properties. 
To demonstrate that, in Fig.~\ref{figure4}(a) 
the dominant gap function (with ${A_{1g}}$ symmetry) is shown for $n=1.0$ and 
$J/U=0.3$. It is clear that this is very similar in 
structure to the {\it subdominant} one shown 
in the previous figures. 
In the inset to Fig.~\ref{figure4}(b) the subdominant gap function
with symmetry ${ B_{2g}}$ is displayed for the same parameters. 
Comparing it with the dominant gap function 
in panel (a) note that, despite having different symmetries, the 
two gap functions are {\it identical} in the two adjacent hole-pocket segments that cross 
the $\Sigma$ line. For this reason, their pairing strengths as measured by $\lambda$ 
(the eigenvalues of the Eliashberg Equation), and shown 
in the main panel of Fig.~\ref{figure4}(b), 
are the same to the third decimal place. Note that the two 
eigenvalues for symmetries ${A_{1g}}$ 
and ${B_{2g}}$ are {\it not} degenerate. This seems a 
strong indication that the ``local'' aspect of the pair scattering, as mentioned above, 
seems to be determinant to establish the pairing properties of this model, at least 
in our RPA weak-coupling approach. It should be noted that the eigenvalue 
results shown in Fig.~\ref{figure4} 
are basically identical to those for lower fillings, 
shown in previous figures, with the only difference 
being the order of the dominant and subdominant symmetries. Since their 
eigenvalues are almost identical for all fillings 
studied, this does not have a special significance. 
Note that $\chi_{\rm RPA}$ for $n=1.0$ and $J/U=0.3$ (not shown) 
follows the same trends as described in Figs.~\ref{figure2} and \ref{figure3}.
From the orbital composition in Fig.~\ref{figure3}(b) and the gap structure in 
Fig.~\ref{figure4} it appears that the symmetry of the B$_{2g}$ and A$_{1g}$ 
pairing operators is determined by the orbitals, while the spatial form in both cases
is characterized by symmetric nearest-neighbor pairing with rotational invariance. Thus, 
the pairing operators have the form 
$\Delta^{\dagger}=f({\bf k})(d^{\dagger}_{{\bf k},X,\uparrow}d^{\dagger}_{-{\bf k},X,\downarrow}\pm d^{\dagger}_{{\bf k},Y,\uparrow}d^{\dagger}_{-{\bf k},Y,\downarrow}$)
where the $+$ ($-$) sign corresponds to A$_{1g}$ (B$_{2g}$) symmetry with $f({\bf k})=~\cos k_x+\cos k_y$, plus higher harmonics with A$_{1g}$ symmetry.

\begin{figure}
\centering
\begin{minipage}{2.0in}
\psfig{file=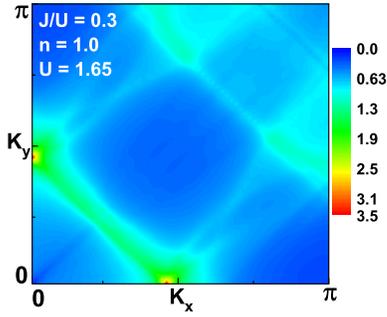,width=2.0in}
\end{minipage}
\caption{(Color online) Two-dimensional plot of the RPA spin susceptibility 
for $n=1.0$. The parameter values are 
$J/U=0.3$ and $U=1.65$. The similarity to the results 
in Fig.~\ref{figure5}(b) is clear, showing also that there are relevant nesting 
features along the $k_y = -k_x + k_{1.0}$ line. Note that a smaller value of $U$ than
in Fig.~\ref{figure5}(a) was used to avoid having a 
peak at $(k_{1.0},0)$ that would wash out the features in the 
rest of the BZ.
}
\label{figure6}
\end{figure}

Figure \ref{figure5}(a) shows in more detail the almost parallel FS segments 
of the two hole-pockets for $n=1.0$. In this figure, the horizontal (blue) vector 
that was defined above as the separation between the two FS segments is displayed. 
A vector with the {\it same} length is reproduced in panel (b), where 
a 2d plot of $\chi_0$ in the first quadrant of 
the BZ is also shown. It clearly indicates that the 
position $\mathbf k_n$ of the main peak in $\chi_0$ is {\it exactly} given 
by the horizontal separation. Not only that, the (red) 
vector along the $\Sigma$ line in panel (a) 
is also reproduced in panel (b) and it coincides also exactly with a 
local maximum of $\chi_0$. In fact (see in both panels 
the black vectors located at angle $\alpha$), 
the locus of the ridge of local maxima in $\chi_0$ in panel (b) exactly coincides with 
the BZ points defined by the vectors connecting the two FS segments 
for $0 \leq \alpha \leq \pi/2$. 
Figure~\ref{figure6} shows the RPA spin susceptibility for $n=1.0$. 
The similarity between these results and those in Fig.~\ref{figure5}(a) is clear,
indicating that the FS nesting for the interacting system is the one described by the 
vectors in Fig.~\ref{figure5}. Finally, an important 
issue should be highlighted: the four points in 
the hole-pockets in Fig.~\ref{figure5}(a) where the gap function has a 
very pronounced peak, are exactly the two pairs of points (one in each pocket) 
connected by $(k_{1.0},0)$ and $(0,k_{1.0})$. This fact clearly links 
the pairing properties with the spin fluctuations. 
Note also that for $n=1.0$ and $J/U=0.3$, 
the second pair of eigenvalues ($\lambda_3=0.9038$ and $\lambda_4=0.9036$) 
corresponds to symmetries ${A_{2g}}$ and ${ B_{1g}}$, 
respectively (not shown). The same occurs for $J/U=0.1$ and $J/U=0.2$, also 
for $n=1.0$ (but the eigenvalues are smaller). Yet, 
the same explanation as described in Fig.~\ref{figure5} applies. 
See the supplemental material for a connection between 
the emergence of a ${B_{1g}}$ symmetry at $n=0.50$
with the one-dimensionality of the bands.

{\it Conclusions.} Summarizing, a weak-coupling RPA analysis 
of a minimal 2-orbital model was used 
to investigate the pairing properties of the BiS$_2$-based 
superconductors. Fillings between $0.46$ 
and $1.0$ were analyzed. The Hund's coupling was varied in the range $0.1 \leq J/U \leq 0.4$. 
Qualitatively, the results are similar for all values of $J/U$ and different fillings. 
In the RPA results described here, a clear relationship 
is found between quasi FS nesting, spin fluctuations, 
and superconductivity: the topology of the two hole-pockets is such that 
they present almost parallel segments close 
to the $(\pi/2,\pi/2)$ wavevector in the BZ. 
It is found that the horizontal distance $(k_n,0)$ between 
the tangents to these segments at the points where 
they cross the $\Sigma$ line is also where the 
non-interacting susceptibility $\chi_0$ has a pronounced 
peak at $(k_n,0)$, for $0.46 \leq n \leq 1.0$. 
Once interactions are introduced, this peak will diverge 
at a certain critical coupling $U$ for each filling, and all 
the values of $J/U$ studied (with exception of one: $n=0.5$, $J/U=0.1$). 
In addition, a line of local maxima, 
connecting the BZ points $(k_n,0)$ and $(0,k_n)$, 
is clearly observed in a 2-d plot of $\chi_0$. As expected, 
this line can also be associated to FS nesting. This nesting 
structure gives origin to pairing functions 
with similar eigenvalues, {\it i.e.}, similar pairing strengths, 
and symmetries ${B_{2g}}$ and ${A_{1g}}$. 
This close competition originates in the FS quasi nesting 
properties, which determine the 
spin-fluctuation-mediated inter-pocket pair scattering. This pair scattering is 
overwhelmingly between two adjacent FS segments, 
therefore the properties of the pairing 
functions, including the pairing strength, are quite ``local'', having almost 
no dependence on their global symmetry. One can then predict that pairing symmetry 
measurements may contain a mixture of both symmetries if the pairing mechanism
is driven by spin fluctuations.  

GBM acknowledges fruitful conversations 
with K. Kuroki, Q. Luo, and H. Usui. 
ED and AM were 
supported by the National Science Foundation Grant No. DMR-1104386.
After finishing this manuscript, a related effort addressing the pairing
symmetry of these materials using a spin model was published. \cite{Hu2012}
There, it is found a dominant ${A_{1g}}$ state analogous to ours, but no competing ${B_{2g}}$
state.

\bibliography{references}

\setcounter{figure}{0}
\makeatletter 
\renewcommand{\thefigure}{S\@arabic\c@figure}

\section{Supplemental Material}

\begin{figure}
\centering
\begin{minipage}{1.6in}
\psfig{file=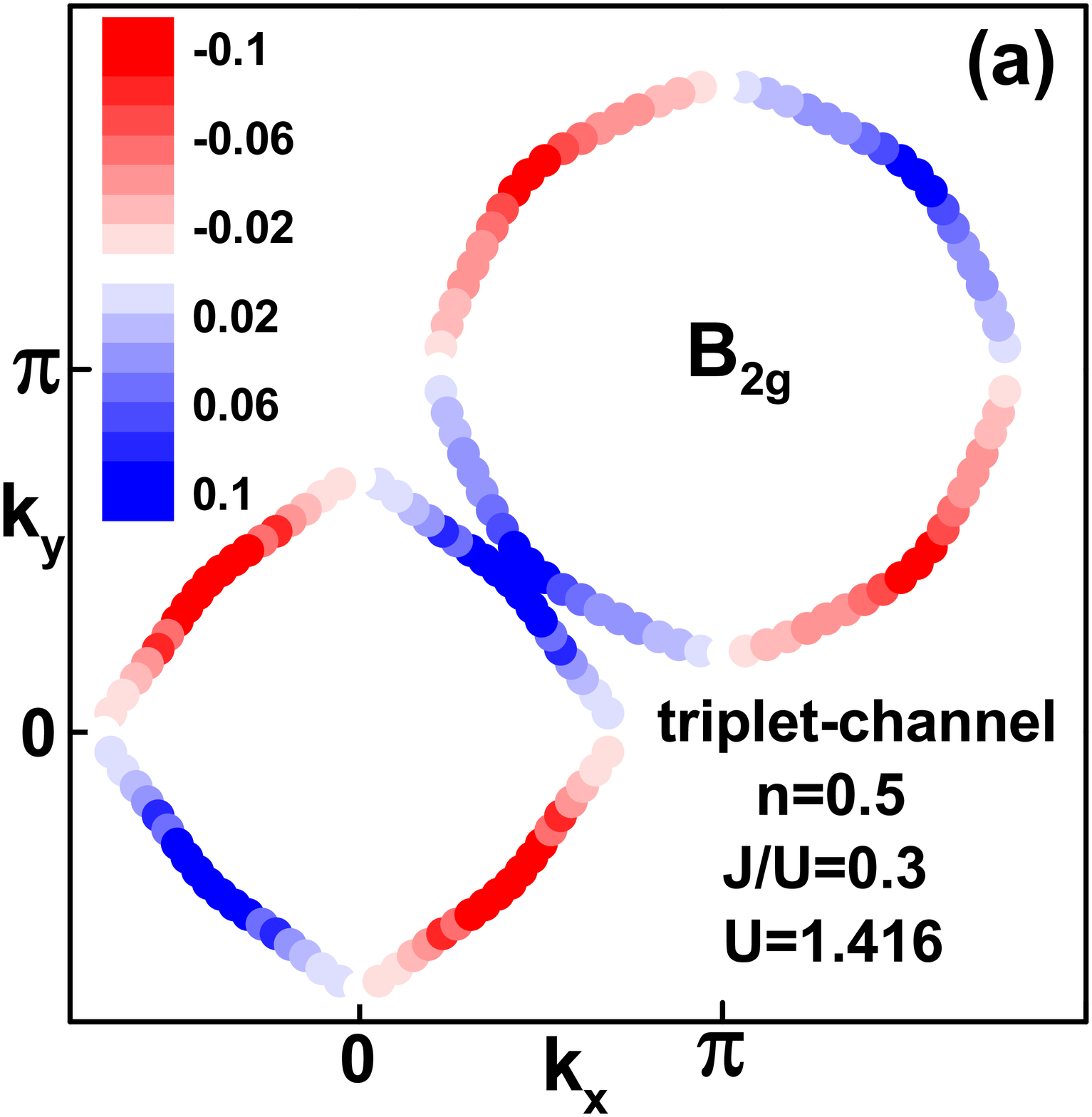,width=1.6in}
\end{minipage}
\begin{minipage}{1.6in}
\psfig{file=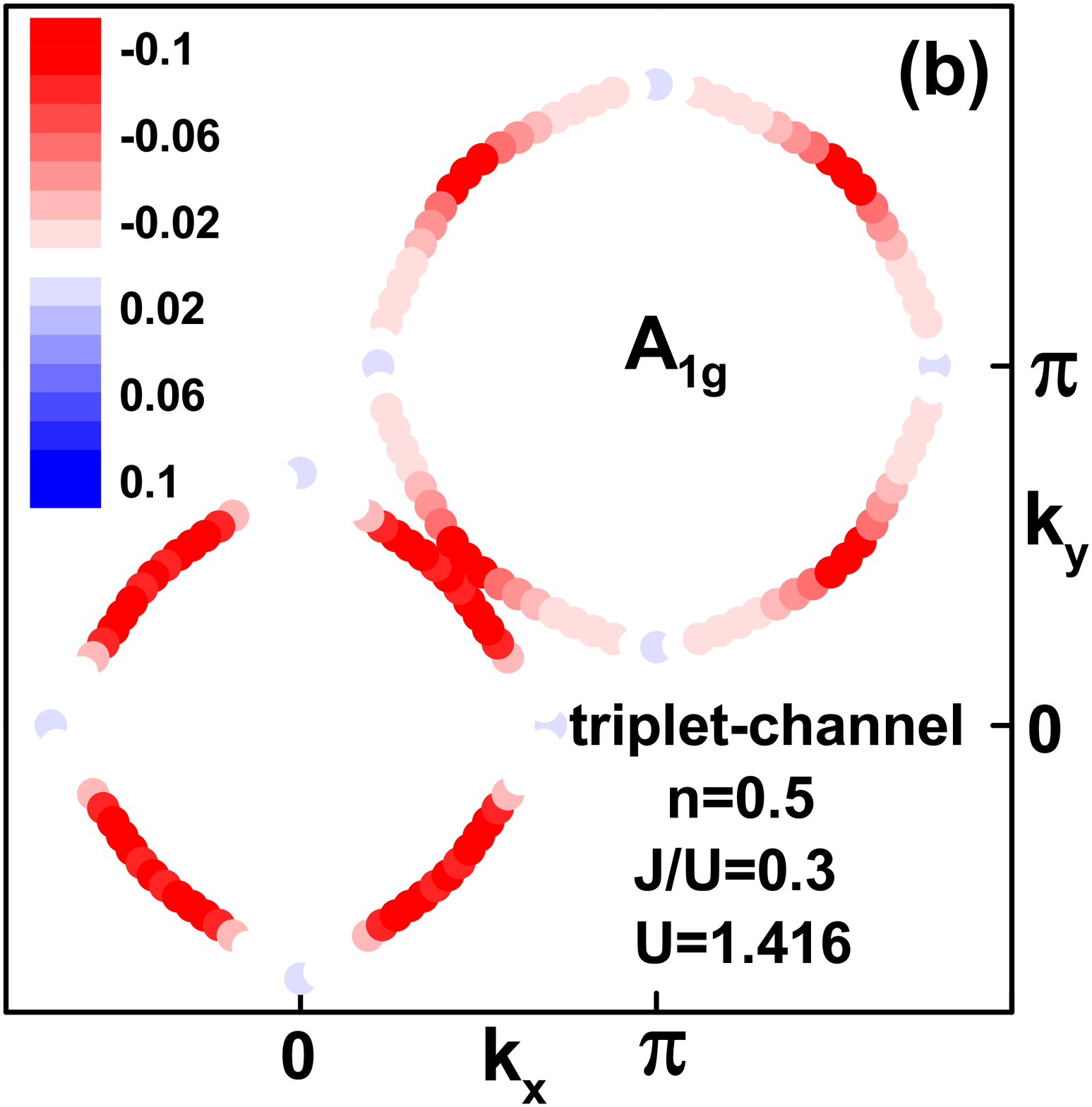,width=1.6in}
\end{minipage}
\caption{(Color online) (a) Dominant and (b) subdominant gap functions 
in the {\it spin-triplet} channel, for 
parameters $n=0.5$, $J/U=0.3$, and $U=1.416$. 
}
\label{figure1s}
\end{figure}

{\it Spin-triplet pairing.}
We also tested the two-orbital model for the case of 
spin-triplet pairing. Using the same RPA 
all the four fillings studied in this work were investigated, but calculations were 
carried out only for $J/U=0.2$ and $0.3$. 
All the critical values obtained for the Hubbard repulsion $U$ were 
slightly above those obtained for the singlet pairing channel. However, they were 
close enough to warrant a brief discussion in this supplemental material. Note that 
in Usui {\it et al.} \cite{Usui2012}
the possibility of spin-triplet pairing was mentioned, 
in connection with the similarity of the ${\rm BiS_2}$ bands 
with those of ${\rm Sr_2RuO_4}$, 
in regards to their common one-dimensionality character. 
Figure \ref{figure1s} shows the 
gap functions [dominant in panel (a) and subdominant in (b)] 
for parameters $n=0.5$, $J/U=0.3$, 
and $U=1.416$. This critical value of $U$ should be 
compared with that obtained for singlet 
pairing for the same parameters ({\it i.e.}, $U=1.363$, see Fig.~\ref{figure2}(b) in the main text). 
The symmetries for 
both the dominant (${B_{2g}}$) 
and subdominant (${A_{1g}}$) gap functions in the 
spin-triplet channel are the same as for the spin-singlet channel. 
The main difference here is that 
they do not have as competing pairing strengths 
as in the spin-singlet channel. Indeed, 
the eigenvalues for Fig.~\ref{figure1s} 
are $\lambda_1=1.00$ and $\lambda_2=0.88$, while 
for the same parameters in the spin-singlet channel their 
values are $\lambda_1=0.989$ and $\lambda_2=0.985$. 

{\it Quasi one-dimensionality.} As mentioned in the main text, the two-orbital model 
has a quasi one-dimensional (1d) character, with the hopping between next-nearest-neighbor being 
dominant ($t_{x \pm y}^{X,Y}=0.88$, see the Table in the main text containing the hoppings). 
It is then interesting to 
verify how the results are modified in case all the other hoppings are removed from the two-orbital model Hamiltonian, 
except for $t_x^{XY}=0.05$. The energies of the orbitals were kept the same as in the original 
model. RPA calculations for the spin-singlet pairing channel were done for $n=0.5$ 
(with corresponding chemical potential $\mu=1.18037278$), $J/U=0.2$, and $J/U=0.3$. 
In addition, the spin-triplet pairing channel was investigated for $J/U=0.3$, but, again, 
the critical value obtained for the Hubbard $U$ was higher than for the singlet channel, 
therefore, these results are not shown. Singlet pairing results for both values of the Hund's coupling 
were similar, therefore, just the results for $J/U=0.3$ will be presented. 

\begin{figure}
\centering
\begin{minipage}{2.5in}
\psfig{file=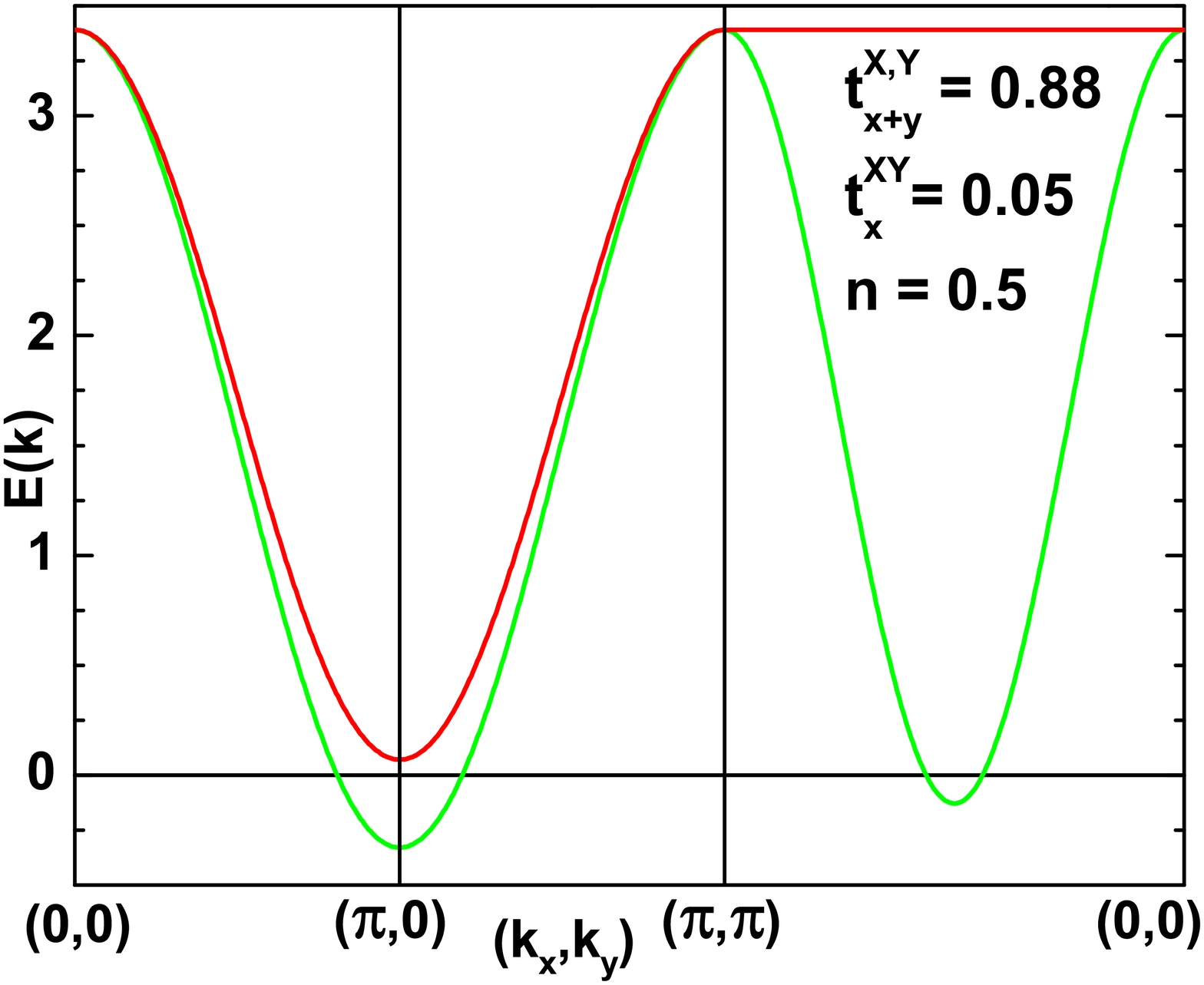,width=2.5in}
\end{minipage}
\caption{(Color online) Band structure for a quasi 1d Hamiltonian obtained from 
the two-orbital model discussed in the main text by keeping only two hopping 
terms: $t_{x \pm y}^{X,Y}$ and $t_x^{XY}$ (see text for details). The Fermi 
energy is at $E_F=0.0$.
}
\label{figure2s}
\end{figure}

Figure \ref{figure2s} shows the band structure for high symmetry lines of the BZ. The Fermi 
energy is located at $E_F=0.0$. The two hole-pockets obtained are identical and nearly square 
(see Fig.~\ref{figure4s}), their corners being slightly rounded due to the presence of the 
finite $t_x^{XY}=0.05$ hopping. For $t_x^{XY}=0.0$ the hole-pockets are perfectly square 
and the two bands are degenerate along the $\Gamma - M$ ($\Sigma$) line.

\begin{figure}
\centering
\begin{minipage}{2.5in}
\vspace{0.5cm}
\psfig{file=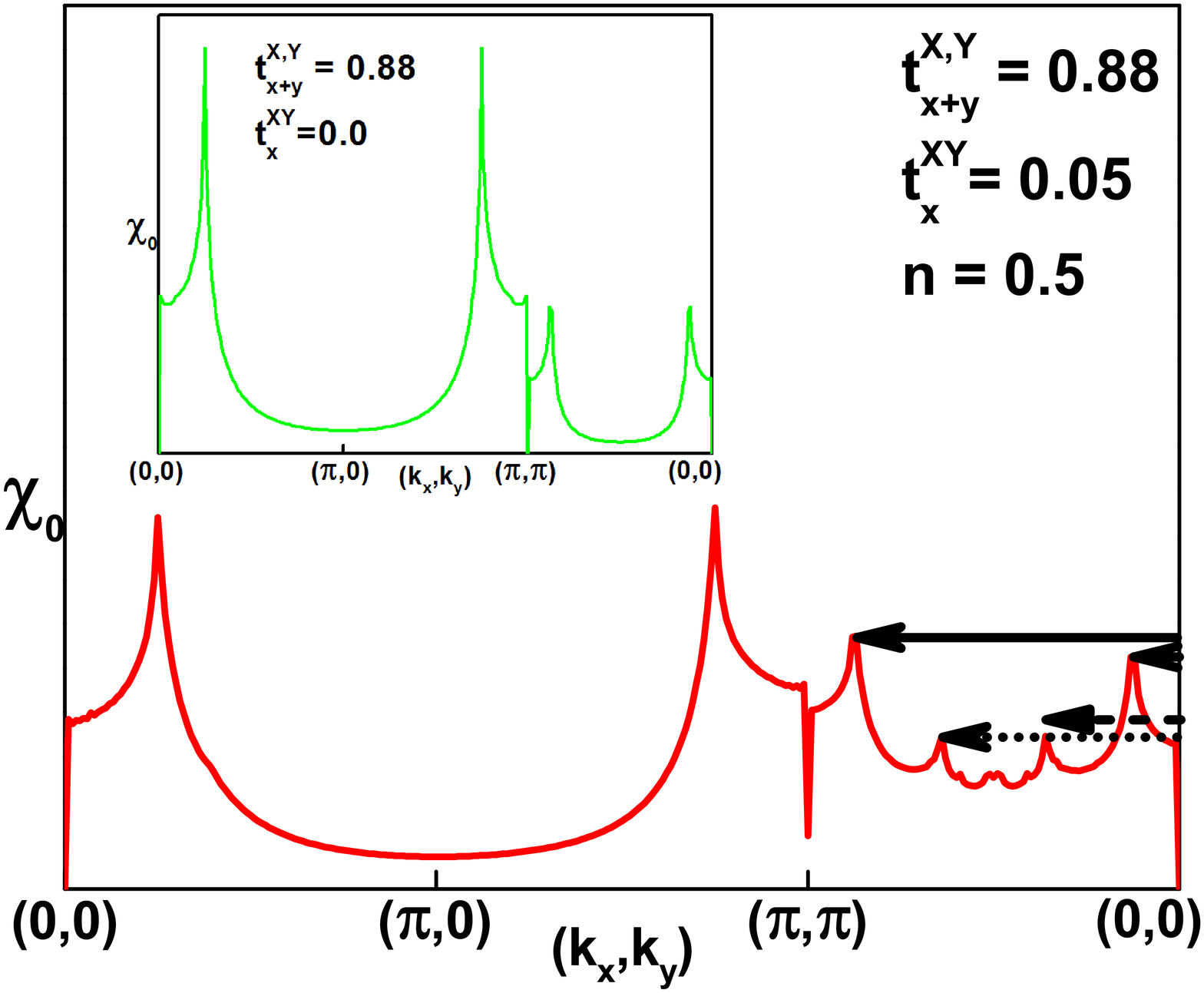,width=2.5in}
\end{minipage}
\caption{(Color online) Main panel: non-interacting magnetic susceptibility $\chi_0$ (Lindhard function) 
for the quasi 1d model. In the inset, $\chi_0$ for the truly 1d Hamiltonian (obtained when 
just the dominant hopping is taken in account, $t_{x \pm y}^{X,Y}=0.88$). The arrows indicating 
peaks located at $(k,k)$ in the BZ are reproduced in Fig.~\ref{figure4s}. These peaks indicate 
different spin fluctuations which may lead to electronic pairing.
}
\label{figure3s}
\end{figure}

\begin{figure}
\centering
\begin{minipage}{1.6in}
\vspace{0.5cm}
\psfig{file=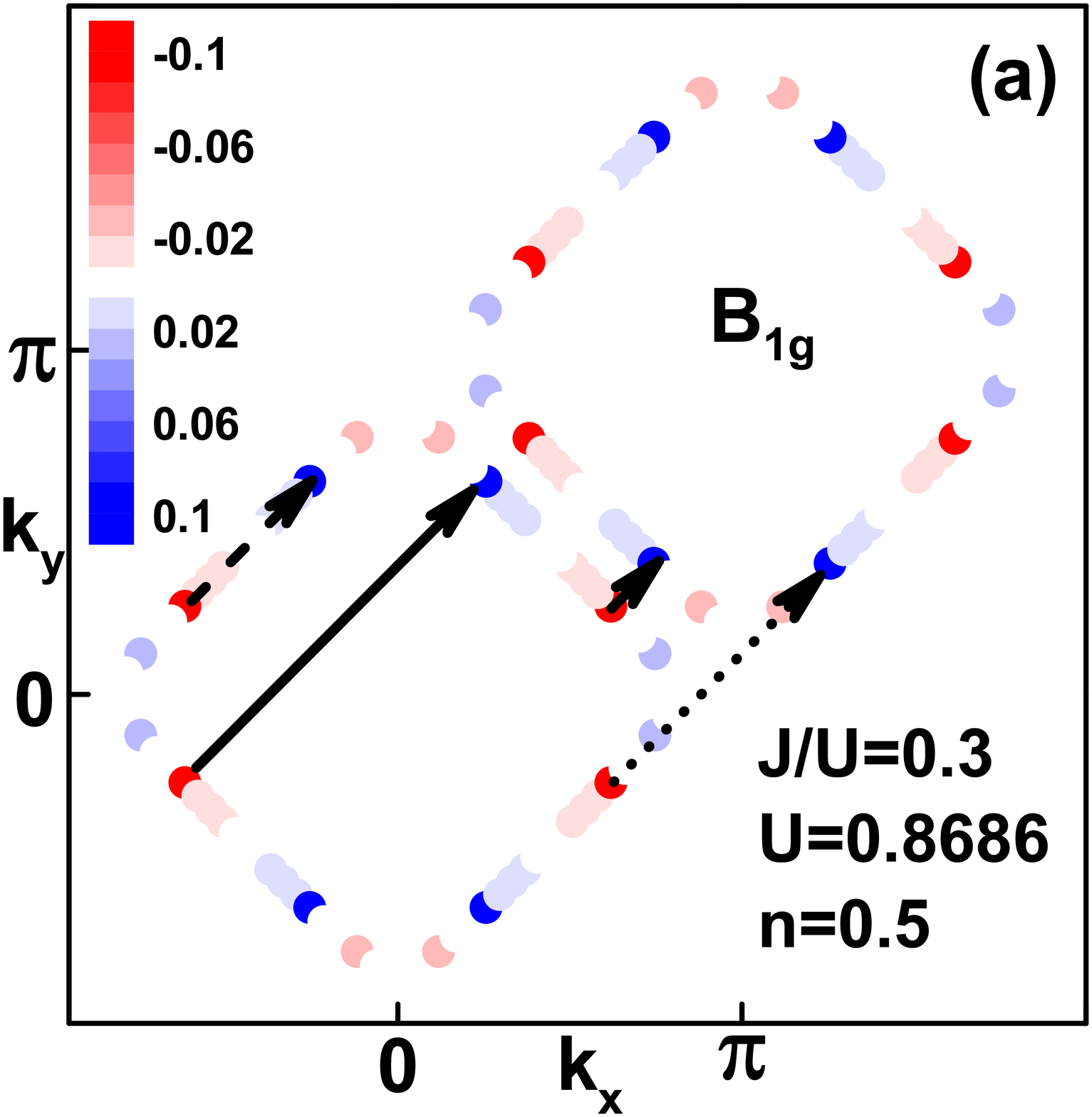,width=1.6in}
\end{minipage}
\begin{minipage}{1.6in}
\vspace{0.5cm}
\psfig{file=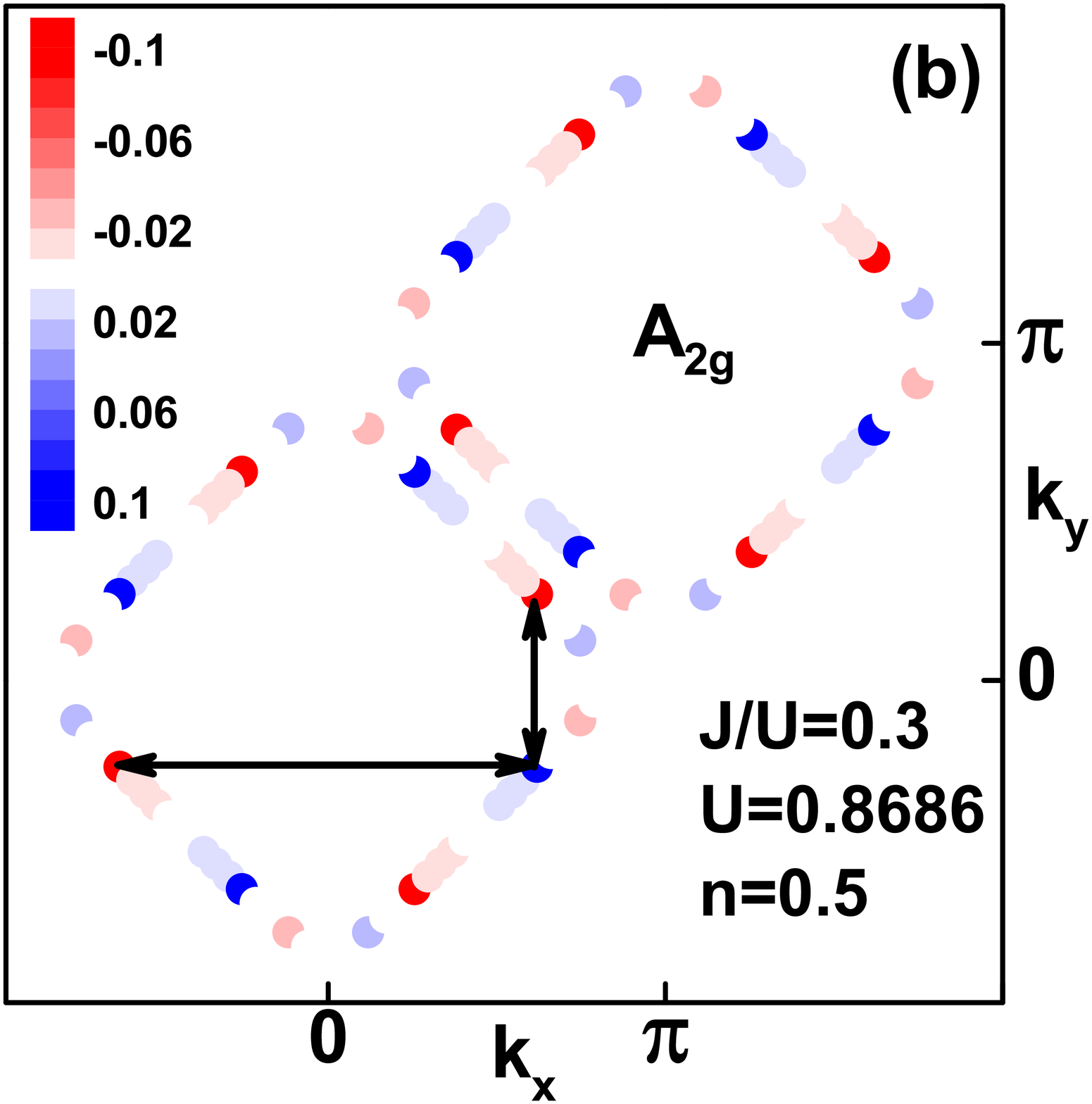,width=1.6in}
\end{minipage}
\caption{(Color online) (a) Dominant and (b) subdominant gap functions 
in the singlet channel for the quasi 1d model. The parameters used are 
$n=0.5$, $J/U=0.3$, and $U=0.8086$. The arrows indicate pair scattering 
processes associated to the peaks in Fig.~\ref{figure3s}, which are 
there indicated with the corresponding line style (see text for details). 
}
\label{figure4s}
\end{figure}

Figure \ref{figure3s} shows, in the main panel (solid red curve), the non-interacting magnetic susceptibility 
(Lindhard function) $\chi_0$ obtained from the bands in Fig.~\ref{figure2s}. The inset shows, as a reference, 
$\chi_0$ for the truly 1d Hamiltonian ({\it i.e.}, $t_x^{XY}=0.0$). A comparison of these 
two $\chi_0$ curves in Fig.~\ref{figure3s} with the one for the fully two-dimensional (2d) two-orbital model (Fig.~\ref{figure1}(b) in the main text) 
shows that the introduction of a small $t_x^{XY}=0.05$ brings the 1d model $\chi_0$ (green curve in the inset) qualitatively 
close to the 2d result. To see that, compare the solid (red) curve in the main pannel of Fig.~\ref{figure3s} with the dashed (green) curve in 
Fig.~\ref{figure1}(b) of the main text. As will be described next, the extra peaks introduced in the $\Sigma$ line (main panel of Fig.~\ref{figure3s}) have a 
marked influence in the singlet pairing gap functions. Indeed, Fig.~\ref{figure4s}(a) shows the dominant gap function 
(with symmetry $B_{1g}$) for $n=0.5$ and $J/U=0.3$. The four vectors connecting local maxima (with opposite signs) 
of the gap function are {\it exactly} the same that locate the four $(k,k)$ peaks in $\chi_0$ in the main 
panel of Fig.~\ref{figure3s}. This, once again, shows the strong connection between spin fluctuations and 
electron pairing. Panel (b) shows the subdominant gap function, with symmetry $A_{2g}$. Their eigenvalues 
are the same up to the third decimal place. It is easy to see that the vectors displayed in the $B_{2g}$ gap function 
[panel (a)] apply identicaly to the subdominant $A_{2g}$ in panel (b). It is also interesting to observe that 
a possible extra set of pair scattering processes, leading to change of sign in the $A_{2g}$ gap function, are 
the ones connecting adjacent sides of the same hole-pocket. Two of them are indicated by double-headed arrows. 
However, these processes do not occur, as there are no peaks in $\chi_0$ that can provide spin fluctuations 
with these two wave vectors (see Fig.~\ref{figure3s}). This results in the pairing strengths of both gap functions 
being basically the same. 

\begin{figure}
\centering
\begin{minipage}{2.5in}
\vspace{0.5cm}
\psfig{file=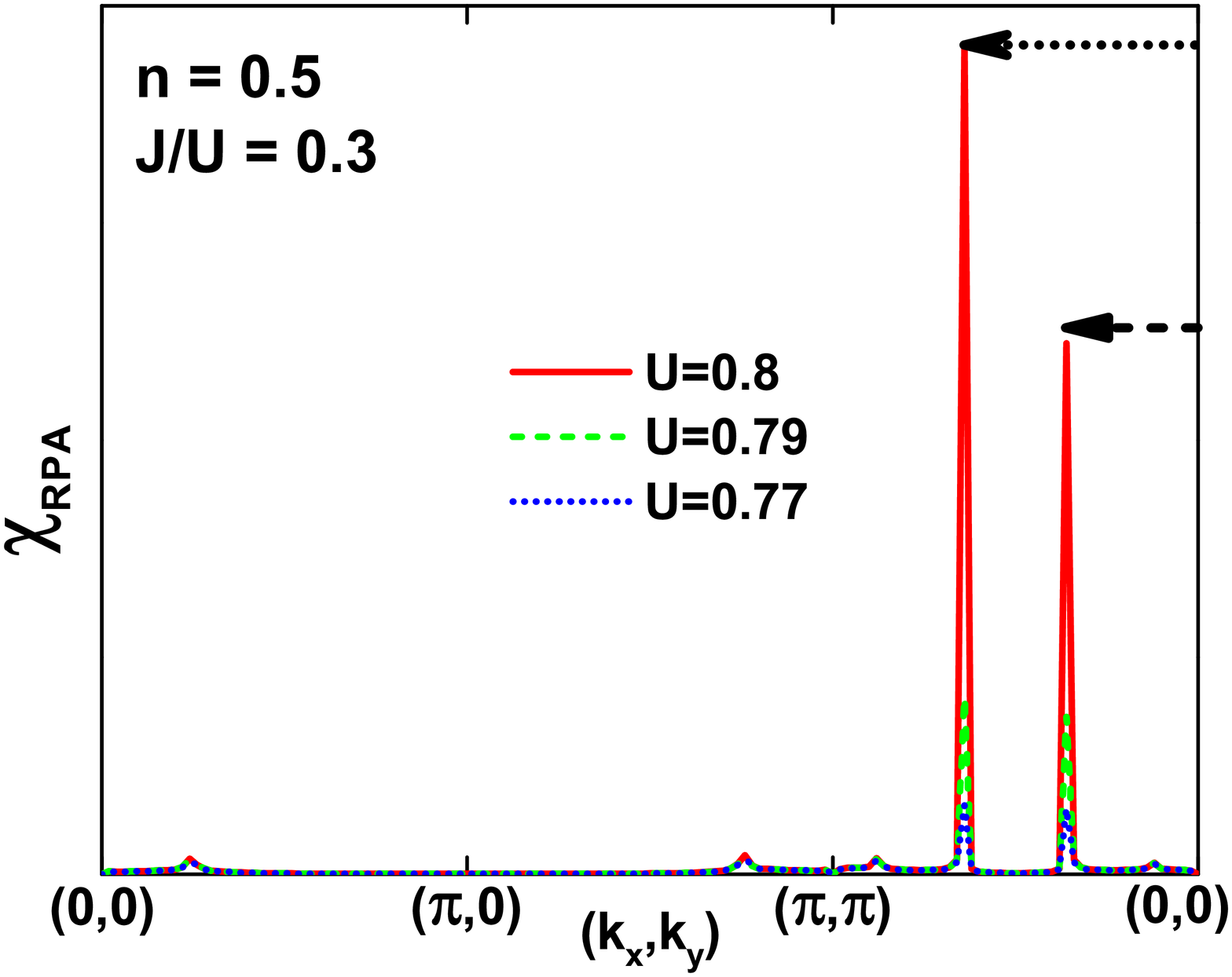,width=2.5in}
\end{minipage}
\caption{(Color online) RPA spin susceptibility for the quasi 1d two-orbital model Hamiltonian, 
$n=0.5$ and $J/U=0.3$. Three curves are shown for values of Hubbard $U$ close to the critical value 
$U_c=0.8686$: $U=0.77$, $0.79$, and $0.8$. A competition between two peaks can be clearly observed. 
They are indicated by the same type of arrows as the ones for the corresponding peaks in Fig.\ref{figure3s}. 
}
\label{figure5s}
\end{figure}

The presence of a relatively large number of different pairing spin fluctuations, as implied in Fig.~\ref{figure4s}, 
suggests that the RPA spin susceptibility should have competing peaks when $U$ is close to the critical value. 
Figure \ref{figure5s} indicates that this is indeed the case. There we show $\chi_{RPA}$ for the same parameters 
as Fig.~\ref{figure4s}, for three different values of Hubbard interaction $U=0.77$ (dotted blue curve), $0.79$ 
(dashed green curve), and $0.8$ (solid red curve). As a comparison, at the same filling $n=0.5$ and $J/U=0.3$, for the full 2d model 
studied in the main text, the leftmost peak (Fig.~\ref{figure2}(b), main text), at a comparable ratio $U/U_c$ 
as the ones in Fig.~\ref{figure5s}, is a few orders of magnitude above the other peaks. Similar results are seen 
for the other fillings and $J/U$ values, indicating that there is mainly a {\it single} dominant pairing 
process in the 2d model in the main text. In the 1d model it seems as if the different wave vector spin-fluctuations 
{\it cooperate} to produce pairing. 

\end{document}